\documentclass{article}
\usepackage[normalem]{ulem}
\usepackage{graphicx}
\usepackage{subfigure}
\usepackage{placeins}
\usepackage{tikz}
\usepackage{dcolumn}
\usepackage{amsmath}
\usepackage{amssymb}
\usepackage{color,colordvi}
\usepackage{float}
\usepackage{epstopdf}
\newcommand{\etal}{\textit{et al.}}
\newcommand{\ts}{\textstyle }
\newcommand{\ds}{\displaystyle}

\def\beq{\begin{equation}}
\def\eeq{\end{equation}}
\def\bea{\begin{array}}
\def\eea{\end{array}}
\def\beqa{\begin{eqnarray}}
\def\eeqa{\end{eqnarray}}

\begin{document}

\title{Semiclassical and quantum description of an ideal Bose gas in a uniform gravitational field}

\author{Rajat K Bhaduri$^1$, Wytse van Dijk$^{1,2}$
%\footnote{\noindent Corresponding author.  \\
%\textit{\mbox{~~~~~}E-mail address:} %vandijk@physics.mcmaster.ca (W. van Dijk)}
}

\maketitle 

%\address
\begin{center}
{$^1$ Department of Physics and Astronomy, McMaster University,   Hamilton, ON, Canada L8S 4M1}

%\address
{$^2$ Department of Physics, Redeemer
University College, Ancaster, ON, Canada L9K 1J4}

%\date{\today}
\end{center}

%\title{Semiclassical and quantum description of an ideal Bose gas in a uniform gravitational field}
%\author{Rajat K. Bhaduri}
%\email{bhaduri@mcmaster.ca}
%\affiliation{Department of Physics and Astronomy, McMaster University,   Hamilton, ON, Canada L8S 4M1}
%\author{Wytse van Dijk}
%\email{vandijk@physics.mcmaster.ca}
%\affiliation{Department of Physics, Redeemer
%University College, Ancaster, ON, Canada L9K 1J4}
%\affiliation{Department of Physics and Astronomy, McMaster University,   Hamilton, ON, Canada L8S 4M1}

\date{\today}

\begin{abstract}
We consider  an ideal Bose gas contained in a cylinder in three spatial dimensions,  
subjected to a uniform gravitational field. It has been claimed by some authors 
that there is discrepancy between the semiclassical and quantum calculations in 
the thermal properties of such a system. To check this claim, we calculate the 
heat capacity and isothermal compressibility of this system semiclassically as well as 
from the quantum spectrum of the density of states. The quantum calculation is done 
for a finite number of particles. We find good agreement between the two calculations 
when the number of particles are taken to be large. 
We also find that this system has the same thermal properties as 
an ideal five dimensional Bose gas.

\end{abstract}

%\maketitle

\vspace{1cm}
\noindent
Keywords:{~~BEC, uniform gravity, isothermal compressibility}

\twocolumn

\section{Introduction}
There is a body of literature on an ideal Bose gas in a uniform gravitational 
field~\cite{gersch57,widom68,baranov01,cavalcanti02,uncu13}.
%,\footnote{For references to early literature, see Ref.~\cite{cavalcanti02}.}.
 The gas may be contained in an external potential, or a large box, and 
subjected to a uniform gravitational potential. Using the semiclassical 
approximation, its thermal properties have been calculated analytically 
in the grand canonical formalism~\cite{liu09a,du12}. Our motivation for studying this 
simple system is to check how closely does the semiclassical approximation follow 
the results of the quantum calculation.  The authors of Ref.~\cite{cavalcanti02} claim that, contrary to previous wisdom, in three dimensions an ideal Bose gas in a uniform gravitational field does not undergo BEC at a finite temperature.    They attribute that to the replacement of the discrete quantum energy spectrum with a smooth density of states.
In the semiclassical approximation, one replaces the discrete density of states by a smooth one, while treating the ground state exactly. In the quantum calculation, on the other hand, the exact discrete energy levels of the system are calculated to compute the grand canonical ensemble (GCE) 
and the resulting thermal properties. In realistic statistical mechanics problems, one generally follows  the semiclassical route. For the system at hand, the quantum 
calculation is done for a finite number of particles. We find that as the number of particles is increased to larger and larger values, the quantum and semiclassical results  become very close, even across the BEC critical temperature.  

In this paper, we pay special 
attention to the calculation of the isothermal compressibility of the Bose gas.   
Recent experimental work on the isothermal compressibility across the Bose-Einstein condensation has been reported in Ref.~\cite{poveda-cuevas15} for a harmonically trapped gas.  The authors suggest that the isothermal compressibility around the critical pressure reveals a second-order nature of the phase transition.  On the other hand predictions based on a number of different mean-field approximations~\cite{olivares-quiroz10} do not lead to second-order phase transitions, and the isothermal compressibility does not diverge at criticality.  In contrast to these authors, we discuss noninteracting systems only.
 It is well documented that in GCE, the isothermal compressibility diverges at the critical temperature $T_c$ in the absence of interparticle interactions~\cite{
%*[{}] [{.~~ See especially, sect. 112 on page 357.}] 
landau58}. 
%and   remains undefined below $T_c$. 
%Since the isothermal 
%compressibility is proportional to number fluctuation in GCE, and below $ T_C$ 
%more and more bosons start occupying the ground state, it is paradoxical that 
%the number fluctuation should diverge. 
It is also known  that even a weak 
interparticle interaction  removes this divergence~\cite{bhaduri02}. In the present 
problem, however, gravitation is  introduced as a one-body ramp potential, and it 
is not clear at the outset how it will affect the compressibility.  
%*[{}] [{,~sect. 12.3, pages 262-272.}] 
%huang63}. 

We find that the semiclassical calculation in three dimensions of the ideal Bose gas with uniform gravity is equivalent to the analysis of a five-dimensional ideal Bose gas without gravity. We use this novel approach to obtain results for the specific heat and isothermal compressibility.   
 The resulting compressibility is divergence-free and continuous across $T_c$.  In the case of heat capacity, 
 in the absence of the gravitational field, there is a discontinuity in its slope at 
$T_c$. Introducing gravitation, or, alternately five spatial dimensions, this  
discontinuity is in the heat capacity itself. 

The calculations were performed by 
taking a cylindrical container, as shown in Fig.~\ref{fig:01}. 
In the $zx$ plane, we take a 
circular disc, which is the bottom of the cylinder at $y=0$. The atoms in the Bose 
gas are not allowed to take negative values of $y$. The gravitational field is along 
the $y$ direction, and the potential is a ramp along the positive $y$ axis.  

The plan of the paper is as follows. In Sec.~\ref{sec:2}, the semiclassical calculation is 
done using the phase space approach. It is established that one can describe the 
system under consideration in five spatial dimensions, but without the gravitation.
 The grand potential is calculated and the critical temperature $T_c$ is obtained. 
In Sec.~\ref{sec:3}, we give the results for isothermal compressibility and the 
heat capacity. In Sec.~\ref{sec:4} a quantum calculation is done to show that BEC takes 
place and the results agree with the semiclassical calculation.  

\section{Three-dimensional  gas in a uniform gravitational potential}         
\label{sec:2}
In this section, we show that an ideal Bose gas in three spatial dimensions,  
subjected to a uniform gravitational potential, may be looked upon as an 
ideal five dimensional gravity-free gas.  We then use the semiclassical method to 
calculate the critical temperature of BEC. 
\begin{figure}[!h]
\begin{center}
\begin{tikzpicture}[scale = 0.5,line width = 1.2pt]
\draw[->](0,0)-- (5,0) node [right]{$x$};
\draw[->](0,0)--(0,5) node [above]{$y$};
\draw[->](0,0)--(-3,-3) node [below,left]{$z$}; 
\draw(0,0) node [left]{$O$};
\draw (0,0) ellipse (3cm and 1cm);
\draw(0,3) ellipse(3cm and 1cm);
\draw(3,0)--(3,3);
\draw(-3,0)--(-3,3);
\draw(3,1.5) node [right]{$L$};
\draw[->](2,4.5)--(1,3) ;
\draw (2,4.5) node [above,right]{Area $=A=\pi a^2$}; 
\end{tikzpicture}
\caption{The Bose gas is confined to a cylindrical box with the downward gravitational force parallel to the $y$ axis.}
\label{fig:01}
\end{center}
\end{figure}
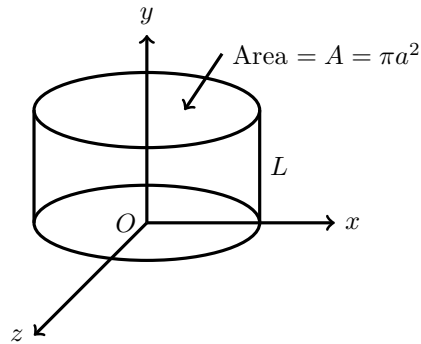

Using the geometry of Fig.~\ref{fig:01}, the single particle energy is given by
\beq
\epsilon (p,y)=\frac {p^2}{2m}+mgy ,
\eeq
where  $m$ is the mass of each boson, and $g$ is the gravitational acceleration 
on the earth's surface, and $p^2=(p_x^2+p_y^2+p_z^2)$. 
 The grand potential is given by ($k_B=1$) 
\beq
\Omega_b=T\sum_n \ln \left (1-z\exp(-\epsilon_n \beta)\right)
=-T\sum_{l=1}^{\infty} \frac{(z)^l}{l} Z_1(l\beta)~,
\label{omega}
\eeq
where $\beta=1/T$, the fugacity $z=\exp(\beta\mu)$, and  
$Z_1(l\beta)$ is the one-body partition function in the variable $l\beta$.
In the semiclassical approximation, $Z_1(\beta)$ in the variable $\beta$, is 
given by 
\beq
Z_1(\beta)=\frac{1} {h^3} \int d^3p~ e^{\ts -\beta p^2/2m} \int d^2 r
\int_0^L dy ~e^{\ts  -\beta mgy}.
\eeq
The two-dimensional spatial integral gives the area $A$ of the disc, yielding  
\beq
Z_1(\beta)=\frac{A}{\lambda_T^3}\frac{(1-\exp{(-\beta mgL)})} {\beta mg}~.
\label{dime1}
\eeq
Note that as $g\rightarrow 0$, we recover the correct 
$Z_1(\beta)=\dfrac{V}{\lambda_T^3}$, 
where $V=A L$ is the three-dimensional spatial volume.  
The thermal wavelength $\lambda_T$ (obtained from the $p$ integration) is given by 
\beq
\lambda_T=\sqrt{\frac{2\pi\hbar^2}{m T}}~.
\eeq
% For  
%our present problem with {\it{nonzero}} $g$, the condition $\beta mgL \gg 1$ must be satisfied in Eq.~(\ref{dime1});  this yields
For our present problem with \textit{nonzero} $g$ and low temperatures, we impose the condition that $k_B T \ll mgL$, i.e. $\beta mgL \gg 1$.  Under this condition, Eq.~(\ref{dime1}) reduces to  
\beq\label{eq:6}
Z_1(\beta)=\frac{A}{\lambda_T^3}\frac{1}{\beta m g}~.
\eeq 
Equation~(\ref{eq:6}) could be rewritten as an ideal five-dimensional partition function (without gravity)
\beq
\tilde{Z}_1(\beta)=\frac{V_5}{\lambda_T^5}
\label{new}
\eeq
where 
\beq
V_5=\frac{2\pi \hbar^2 A}{m^2g}
\label{dime2}
\eeq
has the dimension of (length)$^5$.   We write $V_5=(A\cdot V_3)$, where $V_3$ is a 
hypothetical 3-volume. Taking $m$ to be that of a $Rb^{87}$ atom, we find 
$V_3$ to be exceedingly small, of the order of $10^{-18}$ cubic meter. 
This $V_3$ is not to be confused with the large three-dimensional volume 
$V=AL$ in which the atoms are confined. 
In the following, we shall calculate the thermal properties of this noninteracting gas 
of bosons in 5-spatial dimensions . 
  
Substituting for $\tilde{Z}_1(\beta)$ from Eq.~(\ref{new}) in Eq.~(\ref{omega}), 
we see that the grand potential may be written as 
\beq
\Omega_b=-T \frac{V_5}{\lambda_T^5}\sum_{l=1}^{\infty}\tilde{ b}_l z^l~
=-T \frac{V_5}{\lambda_T^5}~ g_{7/2}(z)~.
\label{cluster}
\eeq
where $\tilde{b}_l=1/l^{7/2}$ are the statistical ``cluster integrals''.  
In standard notation, $\ds \sum_{l=1}^{\infty} z^l/l^{7/2}=g_{7/2}(z) $~.

In the gas phase, 
\beq
\bar{n}_5=\frac{\bar{N}}{V_5}=-\frac{\partial\Omega_b}{\partial\mu}
=\frac{1}{\lambda_T^5
}\sum_{l=1}\tilde{b}_l z^l~=\frac{1}{\lambda_T^5} g_{5/2}(z)~.
\label{deriv}
\eeq
One puts in the constraint that $\bar{N}=N$ , and this makes $z$ a function 
of $T$. The sum on the RHS converges at $z=1$, so the above relation is valid 
only for $T\geq T_c$. For lower temperatures, the ground state starts having 
macroscopic occupancies. The critical temperature is given by 
\beq
(\bar{n}_5\lambda_{T_c}^5)=\zeta(5/2)~,
\label{kya}
\eeq
where $\lambda_T$ is at $T=T_c$, and  $\zeta(5/2)$ is the Riemann zeta function.  
It is straight forward to deduce from Eq.~(\ref{kya}) that the critical temperature 
is given by  
\beq
T_c=\left(\frac{\bar{N}\hbar^3g(2\pi)^{3/2}}{\zeta(5/2)A\sqrt{m}}\right)^{2/5}~.
\label{cond}
\eeq 
This agrees with the expression for $T_c$  as given by Du \etal~\cite{du12}, that was 
obtained by the standard procedure in three spatial dimensions in the 
presence of the uniform gravitational field. Furthermore it follows  that 
\begin{equation}\label{eq:13}
\dfrac{N_{\epsilon=0}}{N} = 1 - \left(\dfrac{T}{T_c}\right)^{5/2} \ \ \ \mathrm{when} \ \ \ T < T_c,
\end{equation}
where $N_{\epsilon=0}$ refers to the number of particles in the ground state.
In Sec.~\ref{sec:4}, a fully 
quantum mechanical calculation is performed to demonstrate that BEC 
does take place at a finite temperature that is consistent with the 
semiclassical result as given by Eq. (\ref{cond}).

\section{Semiclassical isothermal compressibility and heat capacity}
\label{sec:3}
\subsection{Isothermal compressibility }

Quite generally, the isothermal compressibility is defined, in any 
dimension, by 
\beq
\kappa_T=-\frac{1}{V}\left(\frac{\partial V}{\partial P}\right)_T ~,
\eeq
and is directly related to number fluctuation in GCE.
Note that $\kappa_T$ has different dimensionality in three and five 
space dimensions. For this reason, we denote the compressibility in five dimensions 
by $\tilde{\kappa}_T$.

Once the grand potential $\Omega_b$ has been calculated (see Eq.~(\ref{cluster})),
the average particle number $\bar{n}_5$ is obtained from 
$-\dfrac{\partial \Omega_b}{\partial\mu}$  (see Eq.~(\ref{deriv})), 
and its second derivative with respect to $\mu$ is related to $\kappa_T$ :
\beq
-\frac{\partial^2 \Omega_b}{\partial\mu^2}=\frac{\partial \bar{N}}{\partial\mu}
=V_5 \bar{n}_5^2\tilde{\kappa}_T
\eeq
Note that we fix $\bar{N}=N$ using Eq.~(\ref{deriv}), so that the number density of 
bosons remains a constant. 
A little algebra gives, for our five-dimensional ideal gas,
\beq
\tilde{\kappa}_T=\frac{1}{\bar{n}_5^2}\frac{\beta}{\lambda_T^5}{ 
\sum_{l=1}^{\infty}l^2\tilde{b}_lz^l} = \dfrac{1}{\bar{n}_5k_BT}\dfrac{g_{3/2}(z)}{g_{5/2}(z)},
\eeq 
where $\tilde{b}_l=1/l^{7/2}$. Note that $\tilde{\kappa}_T$ is finite and continuous across 
the critical temperature (with $z=1$).   At $T_c$ the compressibility is finite, i.e., $\bar{n}_5k_BT_c\tilde{\kappa}_T=1.9474$. 
This is in contrast to the ideal gas in three spatial dimensions discussed in Ref.~\cite{
%*[{}] [{;~see problem 5 on page 224.}] 
pathria11}.   In that case, the 
compressibility is given as
\begin{equation}\label{eq:16}
\kappa_T = \dfrac{1}{\bar{n}_3k_BT}\dfrac{g_{1/2}(z)}{g_{3/2}(z)},
\end{equation}
where $\bar{n}_3=\dfrac{N}{V} $ is the number density in three dimensions. 
The compressibility  diverges at $T_c$. In the presence of uniform gravitation, 
we find that the above expression is modified to 
\begin{equation}\label{eq:16}
\kappa_T = \dfrac{1}{\bar{n}_3k_BT}\dfrac{g_{3/2}(z)}{g_{5/2}(z)},
\end{equation}
which is not divergent at $T_c$.   Note that this expression is valid when $T\geq T_c$.  

In order to calculate the compressibility below as well as above $T_c$ we use a modification of Eq.~(\ref{deriv}) that includes the effect of the ground state in order to obtain $z$ as a function of $T$, i.e.,
\begin{equation}\label{eq:n3}
\bar{n}_5 = \dfrac{1}{\lambda_T^5}g_{5/2}(z) + \dfrac{1}{V_5}\dfrac{z}{1-z}.
\end{equation} 
The isothermal compressibility is then
\begin{equation}
\label{eq:n20}
\tilde{\kappa}_T = \dfrac{1}{\bar{n}_5k_BT}\dfrac{V_5g_{3/2}(z)+ \lambda_T^5\dfrac{z}{(1-z)^2}}{V_5g_{5/2}(z) + {\lambda_T^5}\dfrac{z}{(1-z)}},
\end{equation}
which diverges at $T=0$.
We use the corresponding equation in three dimensions, with $\kappa_T$ replacing $\tilde{\kappa}_T$, 
$\bar{n}_3$ replacing $\bar{n}_5$, and $V$ replacing $V_5$ in Eq.~(\ref{eq:n20}), to compare with the quantum results in
three dimensions in the presence of uniform gravity.
 As we shall see, this procedure gives good agreement with the quantum calculations for both the heat capacity and the compressibility as shown in Fig.~\ref{fig:06} in  Sec.~\ref{sec:4} of this paper.

\subsection{Heat capacity}
To calculate the heat capacity, we need to calculate the average energy $\bar{E}$, 
which is $-\dfrac{5}{2}\Omega_b$, where $\Omega_b$ is defined in Eq.~(\ref{cluster}), 
and the average number of bosons $\bar{N}$, defined in 
Eq.~(\ref{deriv}).
The energy is differentiated with respect to $T$, with the constraint that 
$\bar{N}=N$, where $N$ is a constant. This gives the condition that 
$\dfrac{d\bar{N}}{dT}=0$, implying that the fugacity $z$ is  dependent on $T$.  
The derivation will not be given here since our result coincides with Du \etal~\cite{du12}. We 
give the final result below, so that this semiclassical result may be plotted 
numerically, and compared with the quantum calculation in the next section. 
\begin{equation}\label{heat}
\def\arraystretch{2.}
\dfrac{C_V}{N} = \left\{ \begin{array}{ll}
\dfrac{35}{4}\dfrac{g_{7/2}(z)}{g_{5/2}(z)}-
                \dfrac{25}{4}\dfrac{g_{5/2}(z)}{g_{3/2}(z)},~~ &T>T_c \\
                       \dfrac{35}{4}\dfrac{\zeta(7/2)}{\zeta(5/2)}\left(\dfrac{T}{T_c}\right)^{5/2},  &T<T_c
                       \end{array} \right.
\end{equation}                       
Noting that for $T\rightarrow \infty$, $z\rightarrow 0$, and in that limit 
$g_{l}(z)\rightarrow z$, we deduce from Eq.~(\ref{heat}) that in this limit 
$\dfrac{C_V}{N}=2.5 k_B$, where we have inserted the Boltzmann constant $k_B$ 
that was hitherto suppressed.  This is in accord with the classical equipartition 
theorem that per degree of freedom the asymptotic heat capacity per particle is 
$k_B/2$.  Therefore the system has five independent dimensions.  Note also the discontinuity at $T_c$, i.e., $[C_V(T_c^{(+)}) - C_V(T_c^{(-)})]/Nk_B = -3.209$. 

\section{Quantum calculation for a system of a finite number of particles}
\label{sec:4}
Quantum calculations for systems with a finite number of particles have been done~\cite{grossmann95b,ketterle96}, which indicate that for trapped particles such systems experience phase transitions even for one or two dimensions.  Such systems have properties that begin to approach even for a small number of particles those in the thermodynamic limit.   Cavalcanti \etal~\cite{cavalcanti02} raise some doubts about the continuous state approximation in the BEC results of the present problem when compared with the quantum calculations.   For that reason we consider  the exact quantum calculation, but with a finite, but increasing, number of particles.

We consider a quantum calculation in three spatial dimensions with the bosons subject to uniform gravity, and compare the quantum results to the semiclassical results in five spatial dimensions.
The energies spectrum is obtained for a particle in a cylindrical space shown in Fig.~\ref{fig:01} with $L$ very large.  In that case, setting $\hbar^2/2m=1$, the energies are
\begin{equation}\label{eq:39}
\epsilon_i = E_{n\nu s} = k^2_{\nu s} +\varepsilon_{n},
\end{equation}
where $\nu$ is the angular quantum number taking values $\nu = 0,\pm 1,\pm 2,\dots$.   The $k_{\nu s}a$ are the zeros  of the Bessel function of the first kind $J_\nu(ka)$,  $\varepsilon_n=-\alpha^{2/3}\eta_n$ where $\eta_n$ are the zeros of the Airy function $\mathrm{Ai}(\eta)$ and $\alpha^2= mg$.  The subscript $i$ refers to all three quantum numbers collectively.  

From the perspective of the bosonic  grand canonical ensemble, the expression for the number of bosons in a system with energy $\epsilon_i$ is
\begin{equation}\label{eq:01}
N_i = \dfrac{g_i}{e^{\ts\beta(\epsilon_i-\mu)}-1} = \dfrac{zg_i}{e^{\ts\beta\epsilon_i}-z},
\end{equation}
where $g_i$ is the degeneracy of the state with energy $\epsilon_i$; $g_i=1$ when $\nu=0$ and $g_i=2$ when $\nu\neq 0$. We use an energy scale such that $\epsilon_0=0$. 
The total number of bosons in the system is
\begin{equation}\label{eq:01a}
%\begin{split}
\bar{N} 
%& =\sum_{i=0}^\infty N_i  = \sum_{i=0}^\infty \dfrac{g_i}{e^{\ts \beta(\epsilon_i - \mu)} - 1} \\ &
= N_0 +  \sum_{i=1}^\infty \dfrac{g_i}{e^{\ts \beta(\epsilon_i - \mu)} - 1}.
%\end{split}
\end{equation}
     To ensure positive $N_i$'s we require that the chemical potential $\mu\leq 0$.  By setting $N_0=0$ and $\mu=0$, the critical temperature $T_c$ can be defined ~\cite{ketterle96}, and we solve
\begin{equation}\label{eq:13a}
\bar{N} = \sum_{i=1}^\infty \dfrac{g_i}{e^{\ts \beta_c\epsilon_i } - 1}
\end{equation}
for $\beta_c$ to obtain $T_c$.

The chemical potential (or the fugacity) is determined by the constraint that the total number of particles is fixed as $N = \bar{N}$.  We can obtain $z$ as a function of $T$ or $T/T_c$ by solving 
\begin{equation}\label{eq:02}
N = \sum_{i=0}^\infty \dfrac{zg_i}{e^{\ts\beta\epsilon_i}-z}
\end{equation}
for $z$ when $T$ is given.  The energy as a function of temperature is
\begin{equation}\label{eq:03}
E = \sum_{i=1}^\infty \dfrac{zg_i\epsilon_i e^{\ts -\beta\epsilon_i}}{1-ze^{\ts -\beta\epsilon_i }}.
\end{equation}
The specific heat is
\begin{equation}\label{eq:35}
\begin{split}
\dfrac{C_V}{Nk_B} & = \dfrac{1}{Nk_B}\left(\dfrac{\partial E}
{\partial T}\right)_{N,V} \\
& = z\left\{\dfrac{\left[ \ds\sum_{i=0}^\infty
 \dfrac{g_i\epsilon_ie^{\ts\beta\epsilon_i}}{(e^{\ts \beta\epsilon_i}-
 z)^2}\right]^2}{\ds\sum_{i=0}^\infty \dfrac{g_ie^{\ts
  \beta\epsilon_i}}{(e^{\ts\beta\epsilon_i}-z)^2}} - \ds\sum_{i=0}^\infty \dfrac{g_i\epsilon_i^2e^{\ts\beta\epsilon_i}}{(e^{\ts\beta\epsilon_i}-z)^2}   \right\},
\end{split}
\end{equation}
where in the derivation we have used the constancy of the particle number, $\dfrac{\ts dN}{\ts dT}=0$, in order to obtain an expression for $\dfrac{\ts\partial z}{\ts\partial T}$.  Finally we obtain the isothermal compressibility 
\begin{equation}\label{eq:38}
\kappa_T = \dfrac{V}{N^2}\left(\dfrac{\partial N}{\partial\mu}\right)_{V,T} = \dfrac{V}{N^2} \beta z \sum_{i=0}^\infty \dfrac{g_ie^{\ts\beta\epsilon_i}}{(e^{\ts\beta\epsilon_i} -z)^2}.
\end{equation} 

\begin{figure}[!h]
\centering                                                                                      
$\begin{array}{c}
\subfigure[~Fugacity as a function of temperature.]{
\resizebox{3.2in}{!}{\includegraphics[angle=-90]
%{z_vs_T_ramp_v2.eps}}
{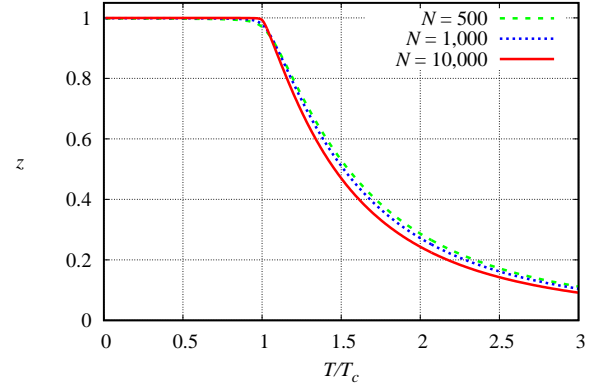}}
\label{fig:05a}}  \\
\subfigure[~Energy-state occupation as a function of temperature.]{
\resizebox{3.2in}{!}{\includegraphics[angle=-90]
%{N_vs_T_ramp_v2.eps}}
{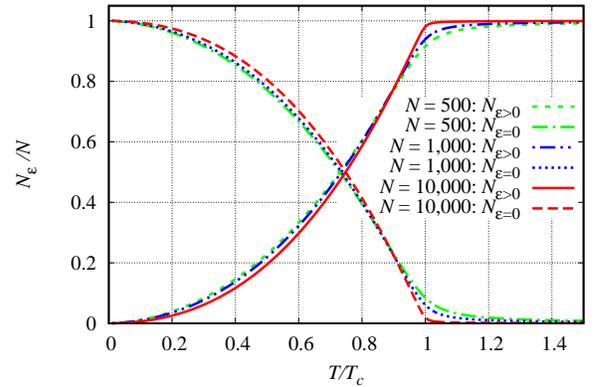}}
\label{fig:05b}}
\end{array}$
\caption{Fugacity and state occupation of system including the ramp potential as a function of temperature.  The results follow from an exact quantum calculation with $a=5$, $\alpha =1$, and $\varepsilon_0=0$.  The curves converge as $N$ increases.}
\label{fig:05}
\end{figure}
\begin{figure}[!t]
\centering                                                                                      
$\begin{array}{c} 
\subfigure[~Specific heat as a function of temperature.  The curves converge to the semiclassical one (in thermodynamic limit) as $N$ increases.]{
\resizebox{3.2in}{!}{\includegraphics[angle=-90]
%{CN_ramp_v2.eps}}
{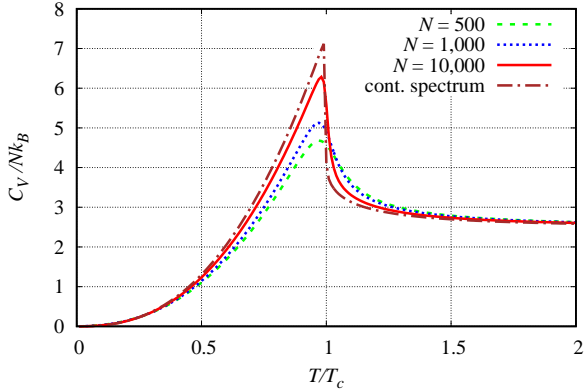}}
\label{fig:06a}}  \\
\subfigure[~Isothermal compressibility as a function of temperature. Note that the vertical axis of the inset is a logarithmic scale.]{
\resizebox{3.2in}{!}{\includegraphics[angle=-90]
%{KT_ramp_v2_log.eps}}
{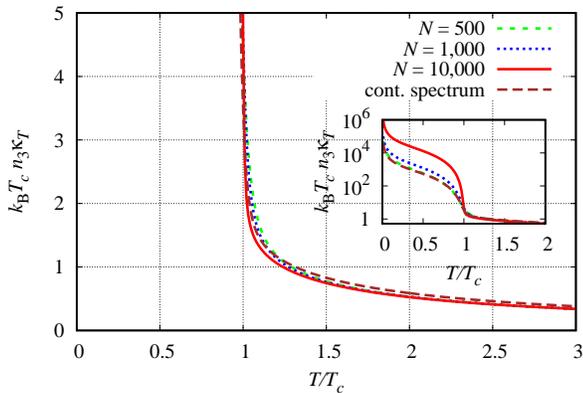}}
\label{fig:06b}}
\end{array}$
\caption{Comparison of quantum and semiclassical calculations of the specific heat and isothermal compressibility.  For the quantum calculation $N= 500, 5,000$ and $10,000$.   Parameters are $a = 5$, $\alpha = 1$, and $\varepsilon_0 =0$.}
\label{fig:06}
\end{figure}

%\FloatBarrier
In Figs.~\ref{fig:05a} and \ref{fig:05b} we show the fugacity and the number of particles distribution as a function of temperature.  In Figs.~\ref{fig:06a} and \ref{fig:06b} we show the specific heat and the isothermal compressibility as functions of the temperature.  In the ramp calculations care must be taken that a sufficient number of energy eigenstate states are included since the energy spacing decreases as the energies increase.  Thus we use 10,000  energy states from the zeros of the Airy functions and we let the maximum values of $\nu$ and $s$ be 200.  A good check to see whether a sufficient number of energy states have been included is the high temperature limit of the specific heat which is 5/2.   The parameters of the calculation are $a=5$ and $\alpha=1$. 

Figures~\ref{fig:05} and \ref{fig:06} clearly show the onset of the BEC; the trend as the particle number increases is already nearly saturated for $N=500$.  The quantum calculations for finite numbers of particles shown in Fig.~\ref{fig:06} approach the semiclassical results which are in thermodynamic limit. The isothermal compressibility has an interesting behaviour  in that the particle number sensitivity seems most pronounced around the critical temperature.  Furthermore the compressibility is finite at the critical temperature and increases rapidly with decreasing temperature.  This corresponds to what is observed in the semi-classical calculation for free bosons in five dimensions.  This is in contrast to the free ideal boson case in three dimensions for which the compressibility approaches infinity when $T$ approaches $T_c$ from above. 
The semiclassical and quantum calculations for $N=500$ are very close.  We used the volume as a parameter in the semiclassical calculation, and it was chosen to be 350 in order that $z$ remain less than or equal to unity.  In all cases we chose $\bar{n}_3=1$.  For consistency with the $N=1,000$ and 10,000 cases we should have chosen $V$ to be 1,000 and 10,000, respectively.  If we do the semiclassical calculations with these values of $V$, the compressibility graphs correspond to the quantum graphs with $N=1,000$ and 10,000, respectively.  In the last two cases the semiclassical values for $z$ slightly exceed unity (by say 0.5\%).
 
%\FloatBarrier

\section{Summary}
In summary, we note that a five-dimensional  gravity-free ideal Bose gas appears to have 
the same thermal properties as that of three-dimensional bosons in a uniform 
gravitational field. The semiclassical approximation using continuous density of 
states is seen to agree with quantum calculations with a finite number of particles, especially as the number of bosons is increased. The divergence in the isothermal compressibility of an ideal Bose gas at $T_c$ is shown to be shifted to zero temperature in  uniform gravity . 

%\section*{References}
%\bibliographystyle{elsarticle-num.bst}
%\bibliography{/home/vandijk/bib_references/ref}

\end{document}